\documentclass[
    ,final            
  ]
  {aipproc}

\layoutstyle{6x9}

\newcommand{\g}{$\gamma$}
\newcommand{\hess}{H.E.S.S.}
\newcommand{\m}{M\,82}
\newcommand{\ngc}{NGC\,253}

\begin{document}

\title{\g-rays from starburst galaxies}

\classification{98.54.Ep} 

\keywords {Galaxies: starburst, Gamma rays: galaxies, Radiation
  mechanisms: non-thermal}

\author{Stefan Ohm} { 
  address={Department of Physics and Astronomy, University of
    Leicester, Leicester, LE1 7RH, UK}, 
  email={stefan.ohm@le.ac.uk},
  altaddress={School of Physics \& Astronomy, University of Leeds,
    Leeds LS2 9JT, UK}}

\begin{abstract}
  In this paper the current status of \g-ray observations of starburst
  galaxies from hundreds of MeV up to TeV energies with space-based
  instruments and ground-based Imaging Atmospheric Cherenkov
  Telescopes (IACTs) is summarised. The properties of the high-energy
  (HE; $100\,{\rm MeV} \leq E \leq 100\,{\rm GeV}$) and
  very-high-energy (VHE; $E \geq100\,{\rm GeV}$) emission of the
  archetypical starburst galaxies \m\ and \ngc\ are discussed and put
  into context with the HE \g-ray emission detected from other
  galaxies that show enhanced star-formation activity such as
  NGC\,4945 and NGC\,1068. Finally, prospects to study the
  star-formation -- \g-ray emission connection from Galactic systems
  to entire galaxies with the forthcoming Cherenkov Telescope Array
  (CTA) are outlined.
\end{abstract}

\maketitle


\section{Introduction}

Starburst galaxies have exceptionally high star-formation rates (SFRs)
in very localised regions (also called starburst region) that have
typical sizes of hundreds of parsecs. In these regions, the high SFR
leads to an increased formation of massive stars and hence supernova
explosion rate compared to the rest of the galaxy. The high thermal
pressure in the central regions often leads to the formation of a
starburst wind. More than 20 years ago, it has been proposed that
starburst galaxies could be sources of HE and VHE \g-ray emission
\citep[e.g.][]{Voelk89,M82:Akyuz91}. The basic idea is that particles
are accelerated in the numerous supernova remnant (SNR) shock waves,
escape into the starburst region and interact with radiation fields,
magnetic fields and with the dense gas. Electrons (and positrons)
predominantly lose energy via inverse Compton and synchrotron
processes, whereas accelerated protons (and heavier nuclei)
predominantly lose energy in proton-proton interactions in which they
produce neutral and charged pions which can decay into e.g. electrons
(positrons) and \g-rays, respectively.

The detection of HE and VHE \g-ray emission from the archetypical
starburst galaxies \m\ and \ngc\ with the \emph{Fermi}-LAT, VERITAS
and \hess\ instruments \citep{Fermi:NGC253M82, VERITAS:M82,
  HESS:NGC253} has triggered a lot of theoretical work. These studies
are aiming to explain the origin and properties of the \g-ray emission
in different scenarios. Most calculations \citep[see
e.g.][]{SB:Lacki11, OhmHinton12, Paglione12, HESS:NGC253_2012}
consider {\it diffuse} \g-ray emission from inelastic proton-proton
interactions and subsequent $\pi^0$-decay as leading loss mechanism,
although it has recently been shown that {\it individual} sources such
as pulsar wind nebulae (PWNe) could significantly contribute to the
\g-ray signal \citep{SB:Mannheim12, SB:Ohm12}.

The first part of this work focuses on the connection between massive
star formation and \g-ray emission in individual Galactic \g-ray
sources and links these to galaxy-size systems such as starburst
galaxies. The spatial and spectral properties of the \g-ray emission
from \ngc\ and \m\ and implications for the origin of the emission are
discussed in more detail in the second part of this paper. In the
third section the \g-ray emission from the starburst galaxies with
Seyfert~2 nuclei NGC\,4945 and NGC\,1068 is addressed. Finally, the
\g-ray emission from other Local Group galaxies is discussed and
prospects for the Cherenkov Telescope Array (CTA) to investigate the
connection between massive star formation and high-energy particles on
all spatial scales is outlined.

\section{Star-formation and \g-ray emission on different spatial
  scales}

At TeV energies, the population of Galactic VHE \g-ray sources
clusters tightly along the Galactic plane. It has a scale height
similar to that of molecular gas and traces regions of massive star
formation. Although $\sim 1/3$ of the source population is still
unidentified and lacks counterparts at lower energies, the majority of
TeV sources is coincident with PWNe, SNRs, SNRs interacting with
molecular clouds or stellar clusters \citep[e.g.][]{HESS:1825,
  HESS:SN1006, MAGIC:W51C, HESS:Wd1_12}. At GeV energies, more and
more of the objects that show bright TeV emission also emerge as HE
\g-ray sources in the \emph{Fermi}-LAT energy band. Especially sources
coincident with regions of active star formation show \g-ray spectra
that smoothly connect from hundreds of MeV to multi-TeV energies. The
most prominent sources, where such spectra are observed are SNRs at
different evolutionary stages that interact with molecular clouds,
e.g. W28 \citep{HESS:W28, Fermi:W28}, W49 \citep{Fermi:W49, HESS:W49},
W51C \citep{Fermi:W51C, MAGIC:W51C} or IC443 \citep{MAGIC:IC443,
  Fermi:IC443}. The preferred interpretation of the \g-ray emission in
these systems is that protons interact with the dense gas and produce
$\pi^0$-decay \g-ray emission.

The progenitors of massive stars that undergo supernova explosions at
the end of their lives typically form and evolve in associations or
stellar clusters. In stellar clusters, the fast winds of massive stars
can merge and/or interact with the SNR shock fronts forming a
collective bubble or even a superbubble. Existing shocks get further
amplified and particles can get accelerated to very high energies
\citep[e.g.][]{Bykov01,Parizot04}. \hess\ observations of the massive
stellar cluster Westerlund~1 revealed an emission region $\sim20$
times the size of the cluster \citep{HESS:Wd1_12}. A very modest
amount of energy injected by the numerous supernovae that exploded
over the past millions of years and the collective stellar winds can
easily explain the emission level. The morphology and spectrum of the
VHE emission suggests that a significant part of the non-thermal
particles originate from Westerlund~1, and that protons are likely
responsible for the emission. At GeV energies, three of the four HE
\g-ray sources that are coincident with the TeV emission are marked as
possibly confused with Galactic diffuse emission, indicating that they
might physically be connected and possibly associated to the \hess\
emission \citep{Fermi:2FGL}. Future studies will show whether or not
the GeV and TeV emission is indeed related and how particle
acceleration, particle escape and interaction with the surrounding
material of Westerlund~1 works in detail. The recent detection of HE
\g-ray emission from the vicinity of the Cygnus superbubble with the
\emph{Fermi}-LAT between 1\,GeV and 100\,GeV further supports the idea
of particle acceleration in stellar clusters and superbubbles
\citep{Fermi:Cygnus}. Also in this case the morphology and spectrum
(that is consistent with an extrapolation to the measurement by
Milagro \citep{Milagro:Cygnus}) suggests that protons are responsible
for the non-thermal emission.

In summary, the study of \g-ray emission from SNRs interacting with
molecular clouds, stellar clusters and superbubbles with Cherenkov
telescopes and the \emph{Fermi}-LAT provided strong evidence for
proton acceleration in regions of massive star formation. All these
Galactic systems show \g-ray emission on spatial scales from tens of
parsecs up to $\sim100$ parsec. The \g-ray emission from starburst
nuclei can be used to investigate the connection between massive star
formation and non-thermal particles on hundreds of parsec scales and
to study populations of particle accelerators. It is complementary to
observations of individual Galactic objects and a powerful tool to
probe the paradigm of cosmic-ray acceleration in SNR shells.

\section{\g-ray observations of starburst galaxies}

The detection of VHE \g-ray emission from the starburst galaxies \ngc\
and M\,82 using ground-based Cherenkov telescopes has been reported in
2009 by the \hess\ \citep{HESS:NGC253} and VERITAS collaborations
\citep{VERITAS:M82}, respectively. Both galaxies are right at the
sensitivity limit of current generation IACTs and it required more
than 100 hours of good-quality data to establish these objects as TeV
\g-ray emitters. \m\ and \ngc\ are the first two external galaxies
that have been detected in VHE \g-rays where the emission is not
powered by a supermassive black hole (as opposed to e.g. Active
Galactic Nuclei or radio galaxies). Shortly after the detection at TeV
energies, the \emph{Fermi}-LAT collaboration announced the discovery
of these two starburst galaxies at GeV energies
\citep{Fermi:NGC253M82}. Also the HE \g-ray signals are very weak and
$\approx 11$ months of LAT data were required to detect \m\ and \ngc\
between a few hundred MeV and a couple of GeV energies. No indication
of variability at any \g-ray energy could be detected from both
starbursts, supporting the idea that the emission is not triggered by
a supermassive black hole at the centres of the galaxies. In the
following the observational properties of the \g-ray emission from
\ngc\ and \m\ are presented and the implications for the origin of the
emission are outlined. A more detailed discussion of theoretical
implications of the \g-ray emission from starburst galaxies can be
found in these proceedings \citep{LackiGamma12}.

\subsection{\ngc}

\begin{figure}\label{fig1}
  \includegraphics[width=0.7\textwidth]{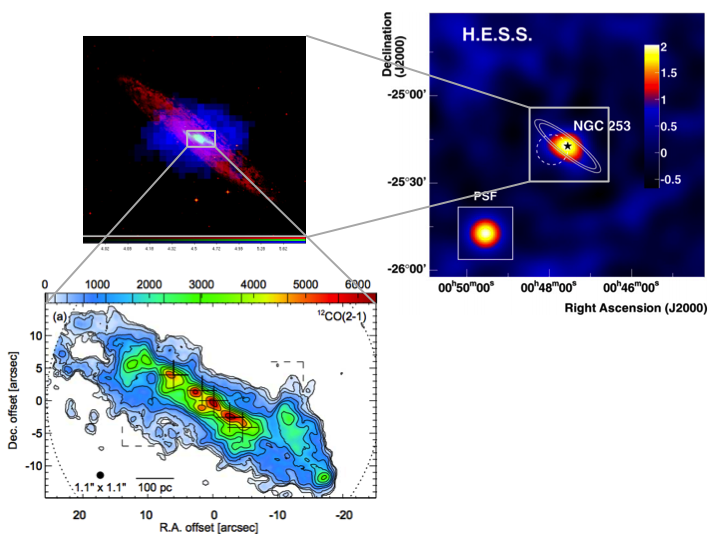}
  \caption{\hess\ sky map of \ngc\ (right,
    \citep{HESS:NGC253_2012}). The zoom into the central part shown at
    the top left is a composite image with optical in red (ESO/DSS),
    near-infrared in green (2MASS, Ks band) and VHE \g-rays in blue
    (H.E.S.S.). The zoom on the bottom left shows the central
    molecular zone of \ngc\ in molecular line emission
    \citep{Sakamoto11} (Reproduced by permission of the AAS).}
\end{figure}

Based on 30 months of LAT data and $\approx 180$ hours of good-quality
\hess\ data, \citep{HESS:NGC253_2012} presented a detailed analysis of
the \g-ray emission from \ngc. The angular extension of \ngc\ of
$\approx 30' \times 7'$ makes it possible to disentangle the TeV
signal from the disc of the galaxy and the starburst nucleus. Indeed,
the upper limit on the extension of the TeV emission of $<2.4'
(3\sigma)$ implies that the VHE \g-ray emission is consistent with
originating in the starburst nucleus. This is also visible in
Figure~\ref{fig1}, where no significant disc component in \ngc\ is
seen at TeV energies. Due to the limited angular resolution of the
LAT, such a discrimination is not possible at GeV energies. The best
fit position of \emph{Fermi} is however consistent with the best fit
position of \hess\ and the optical centre of the galaxy
\citep{HESS:NGC253_2012}.

The large data sets gathered in HE and VHE \g-rays allowed to study
the spectrum of the \g-ray emission in great detail. The \emph{Fermi}
spectrum between 0.2\,GeV and 200\,GeV is best described by a power
law in energy with index $\Gamma_{\rm HE}=2.24 \pm
0.14_{\mathrm{stat}} \pm 0.03_{\mathrm{sys}}$. The \hess\ best fit
spectral index above the energy threshold of 220\,GeV is $\Gamma_{\rm
  VHE}=2.14 \pm 0.18_{\mathrm{stat}} \pm 0.30_{\mathrm{sys}}$ and
hence compatible within statistical errors with the LAT
index. Interestingly, a combined fit of the \emph{Fermi} and \hess\
spectrum over four decades in energy results in a photon index of
$\Gamma=2.34 \pm 0.03$ and a fit probability of 30\%. In case the
emission is of hadronic origin, and assuming canonical values for the
supernova explosion energy $E_{\rm SN} = 10^{51}$\,erg and conversion
efficiency to cosmic rays $E_{\rm SN} \rightarrow E_{\rm CR} = 10\%$,
the measured \g-ray flux suggests that $\approx 20-30$\% of the
protons interact with the dense material in the starburst region
before leaving the system in the starburst wind. Although the
available statistics is quite limited, the smooth alignment of the GeV
and TeV spectrum is suggestive of the fact that one energy loss
mechanism dominates from hundreds of MeV up to several TeV. In this
interpretation energy-dependent diffusion would play a minor role and
(energy-independent) escape of particles from the starburst nucleus
and proton-proton collisions would be the dominant energy loss
mechanisms of the cosmic-ray population in the starburst of \ngc.

\subsection{\m}

\m\ is with an apparent size of $\approx 11' \times 4'$ significantly
smaller than \ngc. It is hence challenging to discriminate the
starburst contribution of the HE and VHE \g-ray emission detected by
the LAT and VERITAS from the disc of the galaxy
\citep{Fermi:NGC253M82, VERITAS:M82}. However, as for \ngc, the GeV
and TeV emission is consistent with originating from a point-like
source located in the starburst core of \m. Based on three years of
LAT data, the \g-ray spectrum of \m\ between 100\,MeV and 100\,GeV can
be described by a power law with spectral index $\Gamma_{\rm HE} = 2.2
\pm 0.1_{\rm stat}$. The measured VERITAS spectrum obtained with a
total of $\approx 140$ hours of good quality data has an index of
$\Gamma_{\rm VHE} = 2.5  \pm 0.6_{\rm stat}  \pm  0.2_{\rm sys}$, and
is within large statistical errors consistent with the \emph{Fermi}
index and extrapolated \emph{Fermi} flux. Within errors, also the flux
from \m\ at GeV and TeV energies is comparable to the flux measured
from \ngc.

\subsection{NGC\,4945 and NGC\,1068}

\begin{figure}\label{fig2}
  \includegraphics[width=0.485\textwidth]{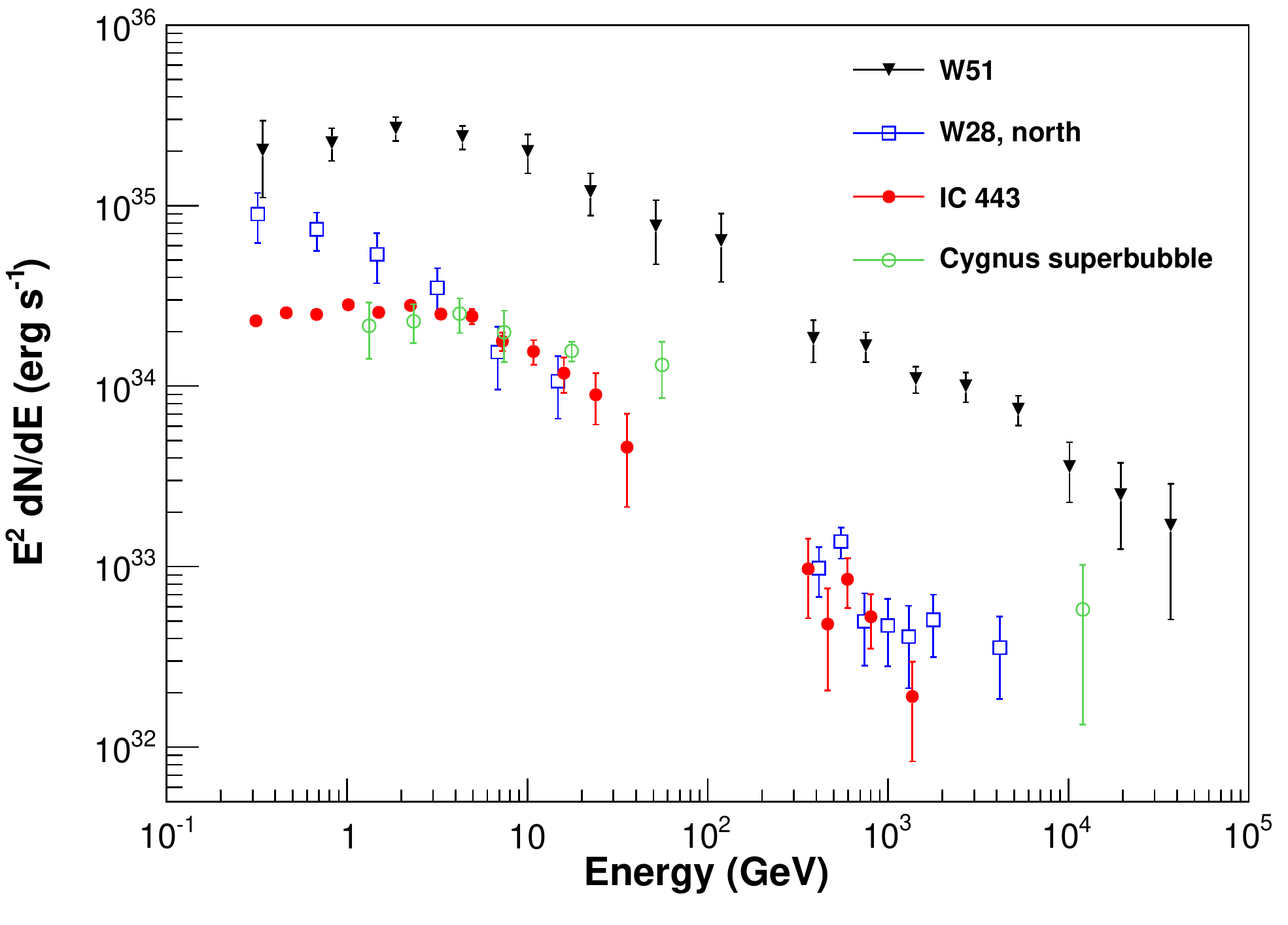}
  \includegraphics[width=0.5\textwidth]{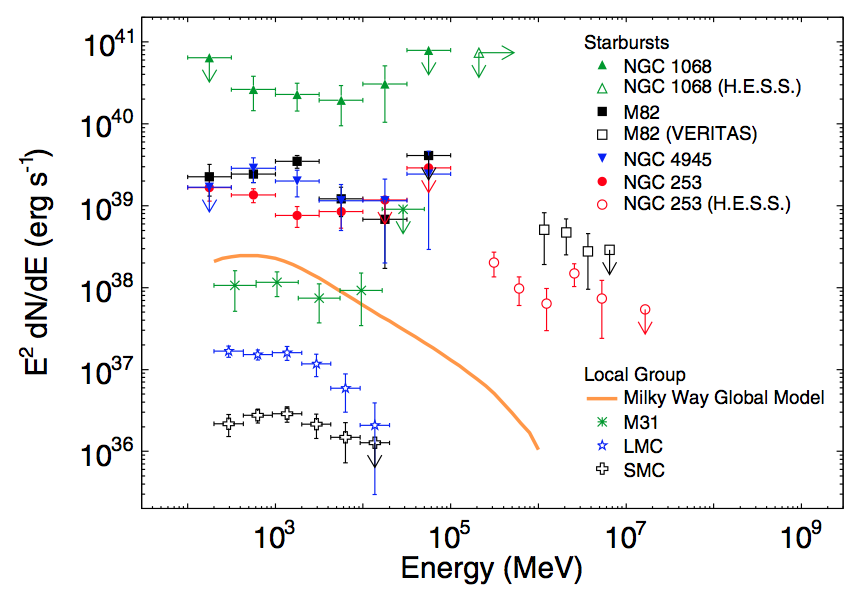}
  \caption{\g-ray luminosity spectra of SNRs interacting with
    molecular clouds and the Cygnus superbubble (left) and of
    star-forming galaxies (right). GeV data has been obtained with
    \emph{Fermi}, distances used to calculate the luminosities are
    given in \citep{Fermi:W28, Fermi:W51C, Fermi:IC443, Fermi:Cygnus,
      Fermi:SB}. TeV data for W28 is from \hess\ \citep{HESS:W28}, for
    W51 from MAGIC \citep{MAGIC:W51C}, for Cygnus from Milagro
    \citep{Milagro:Cygnus} and for IC~443 from VERITAS
    \citep{MAGIC:IC443}.}
\end{figure}

Similar to the two starburst galaxies discussed in the previous
chapter, also NGC\,4945 and NGC\,1068 are galaxies with circumnuclear
starbursts. These two objects however also have radio-quiet obscured
Active Galactic Nuclei (AGN). In 2010, both galaxies were reported to
emit HE \g-rays \citep{Fermi:1yr, Lenain10}. Using three years of LAT
data, \citep{Fermi:SB} analysed the spectrum in more detail and
reconstructed a spectral index of $\Gamma = 2.1 \pm 0.2$ and $\Gamma =
2.2 \pm 0.2$ for NGC\,4945 and NGC\,1068, respectively. The spectra as
shown in Figure~\ref{fig2} (right) are similar to those of \m\ and
\ngc\ and other Local Group galaxies such as M\,31 or the Large
Magellanic Cloud (LMC). So far, no TeV emission from these two Seyfert
2 galaxies has been reported, although a detection is within reach of
current generation IACTs. The HE \g-ray emission from NGC\,4945 and
NGC\,1068 could have their origin in the starburst and/or a possible
AGN activity. One way to disentangle these components is to study the
\g-ray lightcurve, as the starburst emission is expected to be
non-variable. Although statistics is limited, the GeV lightcurves of
NGC\,4945 and NGC\,1068 show no indication for variability, which
could point towards a starburst origin of the \g-ray emission. A
second possibility is to compare the star-formation rate and available
target material in these systems with the observed \g-ray luminosity.
In the case of NGC\,4945, the level of observed \g-ray emission is
consistent with the estimated supernova rate. Together with the
non-variable signal this implies that the starburst activity alone can
account for the observed GeV signal. Although the supernova rate in
NGC\,1068 is comparable to the ones estimated for \m\ and \ngc\ the
observed radio and \g-ray fluxes are an order of magnitude
higher. This suggests that the AGN component might dominate over a
possible starburst activity at these wavelengths. However, apart from
NGC\,4945 and NGC\,1068, no other Seyfert galaxy with radio-quiet AGN
was observed to emit HE \g-rays, with upper limits up to one order of
magnitude below the level seen for NGC\,1068
\citep[e.g.][]{Fermi:Seyferts}. Whether the observed emission from
these two Seyfert galaxies is indeed related to the star-formation
process or if it is rather driven by AGN activity requires further
monitoring by the LAT and search for possible variability as well as
TeV observations and more detailed modelling of the spectral energy
distributions.

\section{\g-ray observations of Local Group galaxies}

Starburst galaxies are not the only galaxies that have been observed
to emit \g-rays. Firstly, the Milky Way itself is a very strong source
of diffuse HE \g-ray emission, originating mainly from cosmic-rays
that illuminate the Galactic interstellar medium and lose energy via
Bremsstrahlung and inverse Compton processes as well as
pion-production and decay. Also the Milky Way satellites, the LMC
\citep{Fermi:LMC} and the Small Magellanic Cloud (SMC,
\citep{Fermi:SMC}) have been detected with the LAT. Moreover, the
Andromeda galaxy (M\,31, \citep{Fermi:M31}) has been reported to emit
GeV \g-rays. The same authors found a simple scaling relation of SFR
and \g-ray luminosity for these Local Group galaxies that also extents
to \ngc\ and \m. \citep{Fermi:SB} used a larger 3-year \emph{Fermi}
data set and examined the \g-ray properties of 69 dwarf, spiral,
luminous and ultra-luminous galaxies in GeV \g-rays. A quasi-linear
scaling relation between \g-ray luminosity and the SFR as traced by
infrared and radio continuum emission further supports the relation
between massive star formation and high-energy
particles. Interestingly, there is a spread of one order of magnitude
in this correlation that could point towards differences in e.g. the
efficiency with which massive stars via e.g. supernovae transfer
energy into non-thermal particles, how high-energy particles interact
with the surrounding material and/or how they escape without
interaction. \citep{Fermi:SB} conclude that between 4\% and 23\% of
the intensity of the isotropic diffuse component as measured with
\emph{Fermi} could be accounted for by unresolved star-forming
galaxies between redshift $0 \leq z \leq 2.5$. Furthermore, up to ten
galaxies might be detectable within ten years of operation.

\section{Prospects for CTA}

Figure~\ref{fig2} (left) shows the \g-ray spectra of three prominent
SNRs that are interacting with molecular clouds and the spectrum of
the superbubble in the Cygnus region. For all these sources, hadronic
emission scenarios are favoured over leptonic scenarios. The same is
true for the \g-ray spectra and the emission from star-forming
galaxies as shown in Figure~\ref{fig2} (right). Especially the smooth
connection of GeV and TeV spectra of {\it individual} Galactic objects
and the {\it diffuse} emission from e.g. \ngc\ and \m\ is striking. So
far, the correlation between star-formation and \g-ray emission has
only be studied for GeV-detected galaxies. At TeV energies, only the
two archetypical starburst galaxies have been detected and no emission
from M\,31, the LMC, SMC or the Milky Way has been reported. This
situation will change once CTA is in operation
\citep{CTA:Actis11}. CTA will deliver an order of magnitude better
sensitivity, a factor $\sim 5$ better point spread function and
broader energy coverage and will allow to study individual objects and
populations of sources in unprecedented detail. First of all, it will
be possible to probe whether or not the \ngc\ TeV emission is
point-like or if it has an extension, similar to the spatial extent of
the central molecular zone in the starburst region. Secondly, it might
be possible to detect the disc of \ngc\ in TeV \g-rays in a deep
exposure, as the expected flux is about one order of magnitude lower
than the starburst emission
\citep[e.g.][]{HESS:NGC253_2012}. Furthermore, the \g-ray spectrum
will be measured with very good accuracy and to lower energies,
allowing to investigate if spectral features arise that indicate a
change in the dominant energy-loss mechanism, or if significant
signatures for \g-\g\ pair production or a dominance of PWNe at TeV
energies become apparent. As for \ngc, also the \m\ \g-ray spectrum
will be studied in great detail and different emission scenarios can
be tested. All other Local Group galaxies that have been detected with
\emph{Fermi} can also be studied with CTA and the SFR -- \g-ray
luminosity relation can be probed with these systems at TeV
energies. Finally, it will be possible to study the whole young
Galactic SNR population with CTA \citep{CTA:SNRs}. With the detailed
understanding of how particle acceleration, escape, and interaction
with the surrounding medium works, it might be possible to get
insights on how the star-formation process is linked to high-energy
particles. The study of superbubbles with CTA might provide the link
between individual \g-ray emitting SNRs and the emission from
starburst galaxies.

\begin{theacknowledgments}
  S.O. acknowledges the support of the Humboldt foundation by a
  Feodor-Lynen research fellowship.
\end{theacknowledgments}

\bibliographystyle{aipproc}

\end{document}